\documentclass[12pt]{article}
\usepackage{graphicx}

\usepackage{amsmath}
\usepackage{gensymb}
\usepackage{floatrow}
\usepackage{caption}
\usepackage{subcaption}

\newcommand\pubnumber{DPF2015-49}
\newcommand\pubdate{\today}

\newcommand{\Eq}[1]{Eq.\ (\ref{eq:#1})}

\def\bnl{Department of Physics\\
Brookhaven National Laboratory, Upton, NY 11973, USA}

\def\Title#1{\begin{center} {\Large #1 } \end{center}}
\def\Author#1{\begin{center}{ \sc #1} \end{center}}
\def\Address#1{\begin{center}{ \it #1} \end{center}}

\newcommand\pubblock{\rightline{\begin{tabular}{l} \pubnumber\\
         \pubdate  \end{tabular}}}
\newenvironment{Abstract}{\begin{quotation}  }{\end{quotation}}
\newenvironment{Presented}{\begin{quotation} \begin{center} 
             PRESENTED AT\end{center}\bigskip 
      \begin{center}\begin{large}}{\end{large}\end{center} \end{quotation}}

\def\beq{\begin{equation}}
\def\eeq#1{\label{#1}\end{equation}}
\def\eeqn{\end{equation}}

\def\beqa{\begin{eqnarray}}
\def\eeqa#1{\label{#1}\end{eqnarray}}
\def\eeqan{\end{eqnarray}}

\let\bar=\overbar

\def\Dslash{\not{\hbox{\kern-4pt $D$}}}
\def\dslash{\not{\hbox{\kern-2pt $\del$}}}

\def\msb{{\bar{\ssstyle M \kern -1pt S}}}

\pdfoutput=1

\begin{document}
\begin{titlepage}
\pubblock

\vfill
\Title{Signal Processing in the MicroBooNE LArTPC}
\vfill
\Author{ Jyoti Joshi, Xin Qian \\
(For THE MicroBooNE Collaboration) 
}
\Address{\bnl}
\vfill
\begin{Abstract}
The MicroBooNE experiment is designed to observe interactions of neutrinos with a 
Liquid Argon Time Projection Chamber (LArTPC) detector from the
on-axis Booster Neutrino Beam (BNB) and off-axis Neutrinos at the Main
Injector (NuMI) beam at Fermi National 
Accelerator Laboratory. The detector consists of a $2.5~m\times 2.3~m\times 10.4~m$ TPC including an array 
of 32 PMTs used for triggering and timing purposes. The TPC is housed in an evacuable and foam insulated cryostat 
vessel. It has a 2.5 m drift length in a uniform field up to 500 V/cm. There are 3 readout wire planes 
(U, V and Y co-ordinates) with a 3-mm wire pitch for a total of 8,256 signal channels. The fiducial mass of the 
detector is 60 metric tons of LAr.\\

In a LArTPC, ionization electrons from a charged particle track drift  along the electric 
field lines to the detection wire planes inducing bipolar signals on the U and V (induction) planes, and a unipolar 
signal collected on the (collection) Y plane. The raw wire signals are
processed by specialized low-noise front-end readout electronics immersed in LAr which
shape and amplify the signal. Further signal processing and
digitization is carried out by warm electronics. 
We present the techniques by which the observed final digitized waveforms, 
which comprise the original ionization signal convoluted with detector
field response and electronics response as well as noise, are processed to recover 
the original ionization signal in charge and time.
The correct modeling of these ingredients is critical for further event reconstruction in LArTPCs.
\end{Abstract}
\vfill
\begin{Presented}

DPF 2015\\
The Meeting of the American Physical Society\\
Division of Particles and Fields\\
Ann Arbor, Michigan, August 4--8, 2015\\
\end{Presented}
\vfill
\end{titlepage}
\def\thefootnote{\fnsymbol{footnote}}
\setcounter{footnote}{0}

\section{Liquid Argon Time Projection Chambers}
Liquid Argon Time Projection Chambers (LArTPCs)~\cite{lartpc_1, lartpc_2} 
provide a powerful, robust, and elegant solution for studying neutrino interactions and 
probing the parameters that characterize neutrino oscillations. LArTPC technology offers a 
unique combination of millimeter scale 3D precision particle tracking and calorimetry with 
good dE/dx resolution. This combination results in high efficiencies for particle 
identification and the background rejection. Due to its scalability and fine grained tracking
capability, LArTPC technology is a promising choice for the next generation massive neutrino 
detectors. Liquid Argon is an ideal medium since it has high density, excellent properties 
such as large ionization and scintillation yields, is intrinsically safe and cheap, and is readily 
available anywhere as a standard by-product of the liquefaction of air. The operating principle 
of large scale LArTPC detectors is based on the fact that in highly purified liquid argon, ionization tracks can 
be transported by a uniform electric field over distances of the order of meters.

\begin{figure}[htb]
\centering
\includegraphics[height=3.0in]{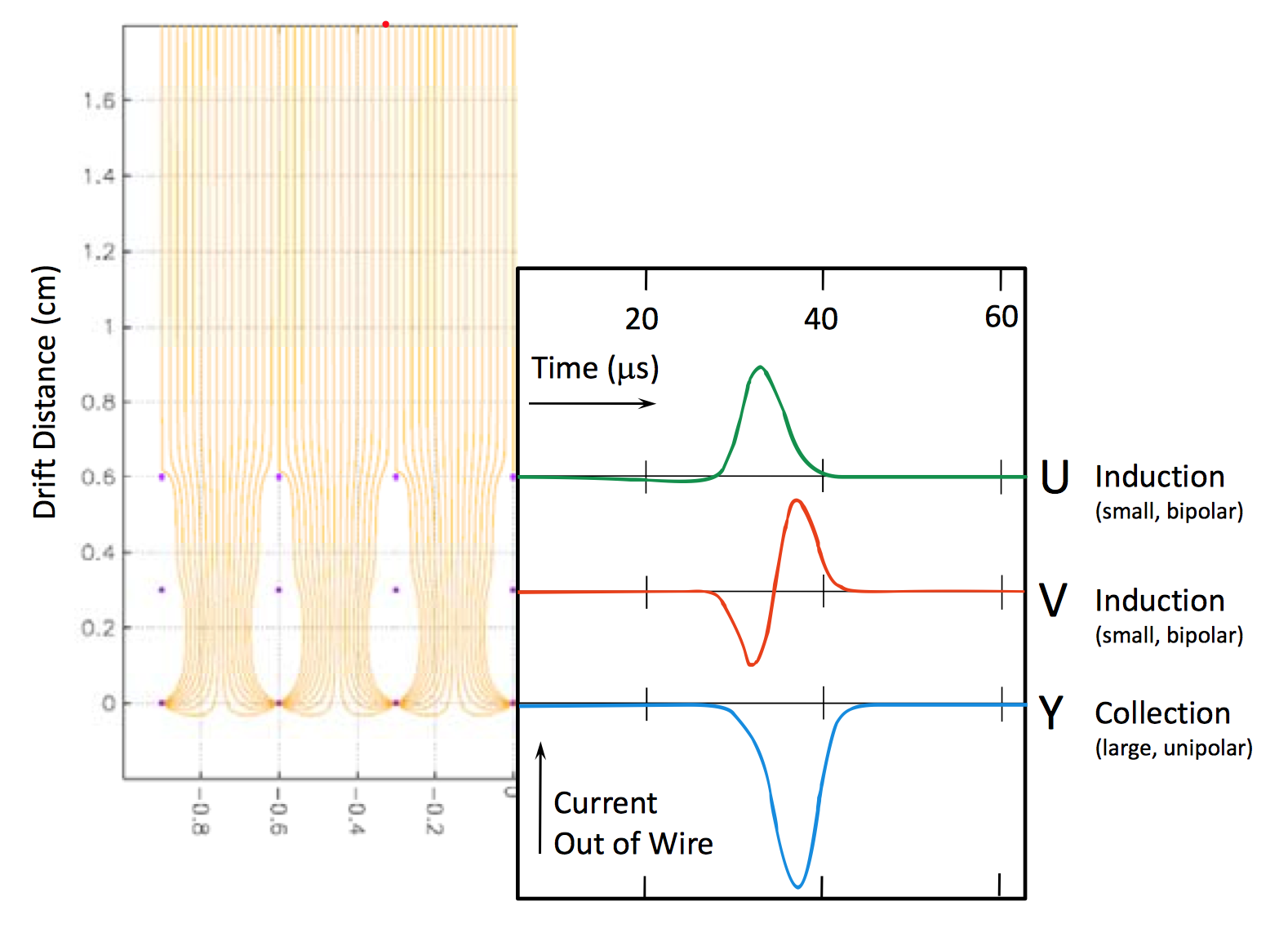}
\caption{The signal properties of LArTPC}
\label{fig:signal}
\end{figure}

A single-phase LArTPC is basically a tracking wire chamber placed in
highly purified liquid argon with an electric field created within the
detector. Ionization electrons produced when charged particles go through 
the detector volume would drift along the electric field until they reach the wire-planes and 
hence produce signals that are utilized for imaging purposes. Several wire-planes 
with different orientations using  bias voltages chosen for optimal field shaping give several complimentary views of the same 
interaction as a function of drift time, providing the necessary information for reconstructing a
three-dimensional image of the interaction~\cite{wireplanes}.\\

A schematic illustrating the LArTPC signal response is shown in Fig.~\ref{fig:signal}. It shows the 
planar illustration of electric field lines (i.e, electron trajectories) and the signals 
induced by an ionizing track at $90\degree$ to the wire direction and at $0\degree$ to the 
wire planes. In the simulation, the wires in the induction planes U and Y are inclined at 
$\pm45\degree$ with respect to wires in the Y collection plane. Bipolar signals from two induction 
planes, and the unipolar signal from the collection plane are
processed and readout by specialized low-noise front-end readout electronics immersed in LAr. 

\section{The MicroBooNE LArTPC}
MicroBooNE~\cite{uboone_1, uboone_2} is newly built LArTPC neutrino detector of 60 metric ton 
fiducial mass (170 ton total) at Fermilab National Accelerator
Laboratory in Batavia, Illinois. MicroBooNE recently started its 
operation and has been collecting neutrino data from the Booster
Neutrino Beamline (BNB) since October, 2015. 
The experiment's primary motivation is to resolve the source of the
MiniBooNE low energy excess observed in $\nu_e$ candidates by taking advantage of the excellent electron-photon
particle identification capabilities of a LArTPC in addition to
carrying out a comprehensive suite of neutrino cross section measurements on Argon. 

\begin{figure}[htb]
\centering
\includegraphics[height=3.0in]{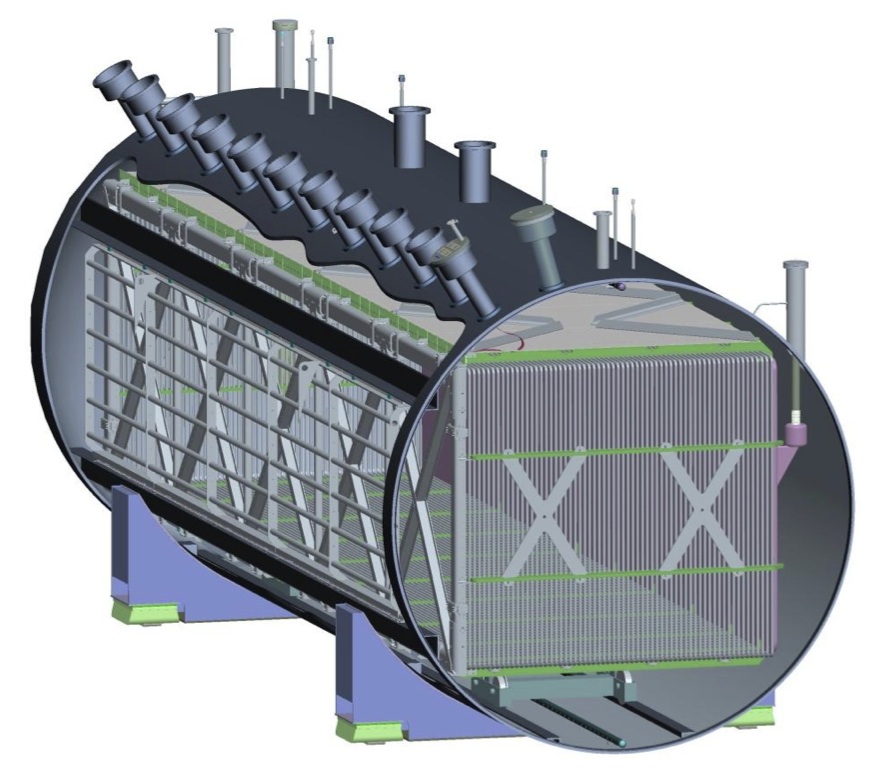}
\caption{Schematic diagram of MicroBooNE detector}
\label{fig:uboone}
\end{figure}

The TPC itself contains three wire planes, one collection plane at $0\degree$ from vertical and two 
induction planes at $\pm60\degree$ with 3-mm wire pitch and 3-mm wire
plane separation. MicroBooNE also serves as a test bed for LArTPC 
technologies for next generation very large scale detectors. There are many innovative technologies 
implemented, such as 2.5 m long drift distance, cold front-end
low-noise readout
electronics~\cite{cold} and a filling procedure 
that does not include prior evacuation of the cryostat while still
maintaining ultra-high purity LAr. The use of cold electronics within
the LAr volume is critical for enabling the scaling up of the LArTPC
technology and to improve the signal-to-noise ratio. The light collection system~\cite{pmt} consists of 32 8-inch PMTs that are located just behind the wire planes to detect scintillation light from $\nu$-Ar interactions. The PMT information is used to trigger 
on beam events and significantly reduce the data throughput. A
schematic diagram of the MicroBooNE detector is shown in Fig.~\ref{fig:uboone}.

\section{Signal Processing Chain}
The raw signal on the wires in the TPC consists of both the ionization
signal and the noise. The signal 
is a convolution of the distribution of the electron cloud passing through the TPC wires, the field response 
(i.e. the induced current on wires), and the electronics response. The background includes the noise 
from various sources. The goal of signal processing is to extract both
the signal charge and time information 
reliably and separate it from the noise. The following subsections will give details on each step.

\subsection{Field and Electronics Response Modeling}
\begin{figure}[!h!tbp]
\includegraphics[width=0.48\textwidth]{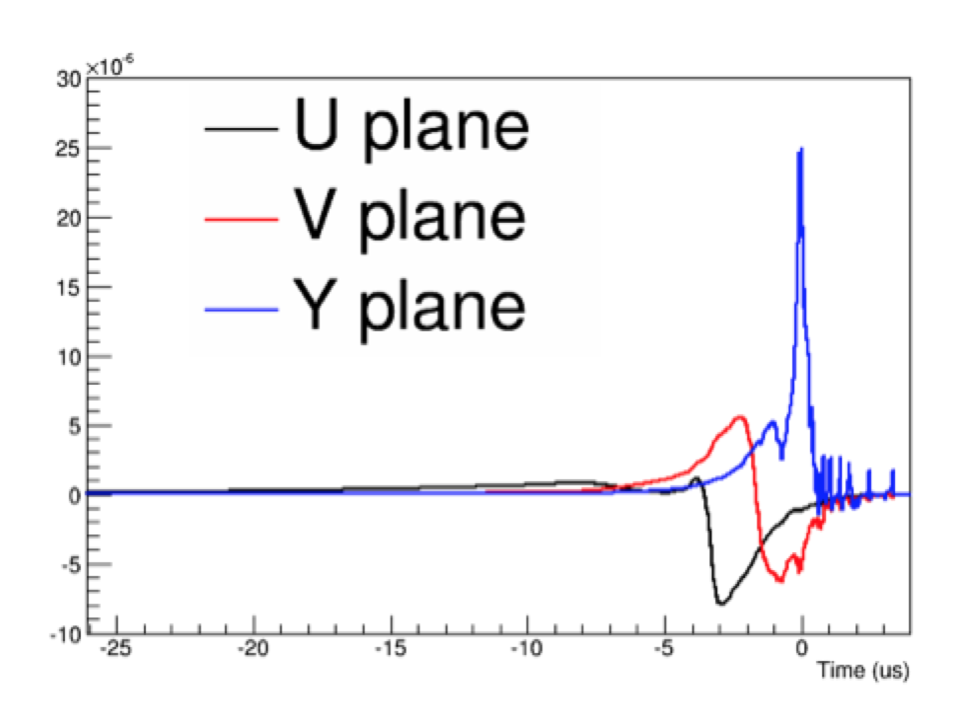}
\includegraphics[width=0.48\textwidth]{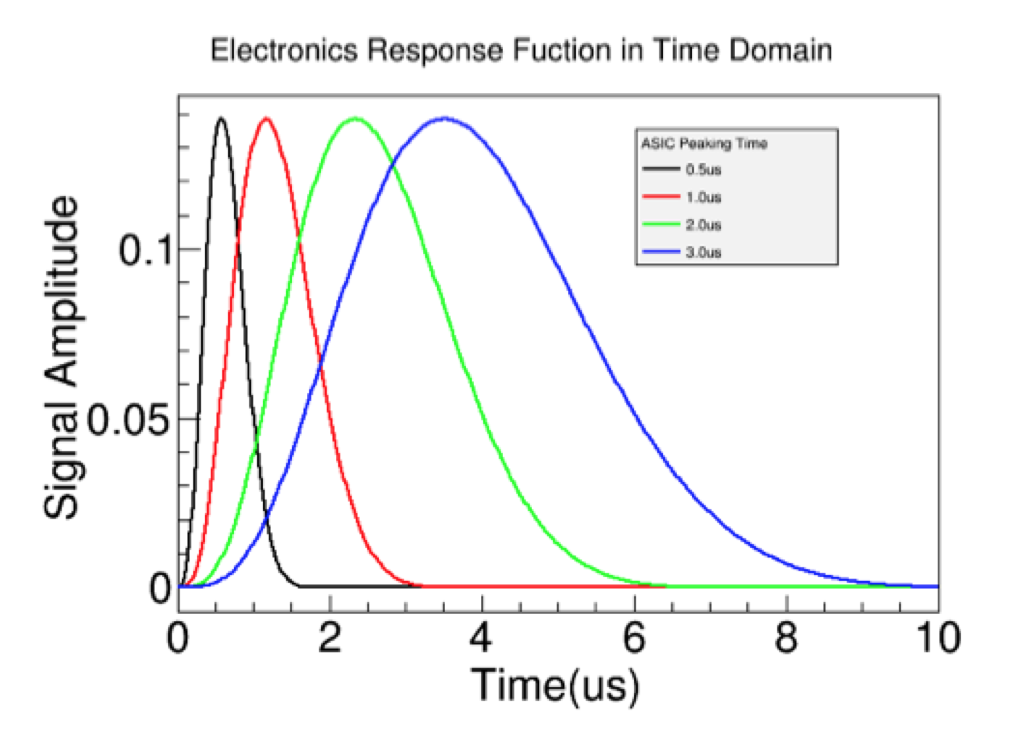}
\caption[resp]{Field Response Functions (left) and Electronics Shaping Functions (right)}
\label{resp}
\end{figure}

A detailed knowledge of the field and electronics response is necessary in order to characterize the 
detector performance. Simulating the field response function is the first step in the chain of signal processing. 
The drifting electrons are modeled as many small clouds of charge that diffuse as they travel toward 
the collection wires. The response of the channels to the drifting electrons is parameterized as a function 
of drift time, with separate response functions for collection and induction wires. The signals on the 
induction-plane wires result from induced currents and are thus bipolar as a function of time as charge 
drifts past the wires, while the signals on the collection plane wires are unipolar.
Fig.~\ref{resp} (left) shows the 2-D GARFIELD~\cite{garfield}
simulated response to a single electron blob generated in the MicroBooNE detector
geometry in terms of charge vs. time averaged for a single electron for both induction planes (U-Plane 
in black and V-Plane in red) and collection plane (in blue). 

The electronics response function for the MicroBooNE detector is shown in Fig.~\ref{resp} (right) in 
terms of signal amplitude vs. time. Since the MicroBooNE front-end cold
electronics are designed to be programmable with 4 different 
gain settings (4.7, 7.8, 14, and 25 mV/fC) and 4 shaping time settings
(0.5, 1, 2, and 3 us), the electronic 
response function varies according to these settings. Different colored lines in Fig.~\ref{resp} (right) 
show the electronics response for different shaping time settings. For a fixed gain setting, the peak 
is always at the same height independent of the shaping time.

\subsection{Raw and Convoluted Signal}
\begin{figure}[!h!tbp]
\includegraphics[width=0.32\textwidth]{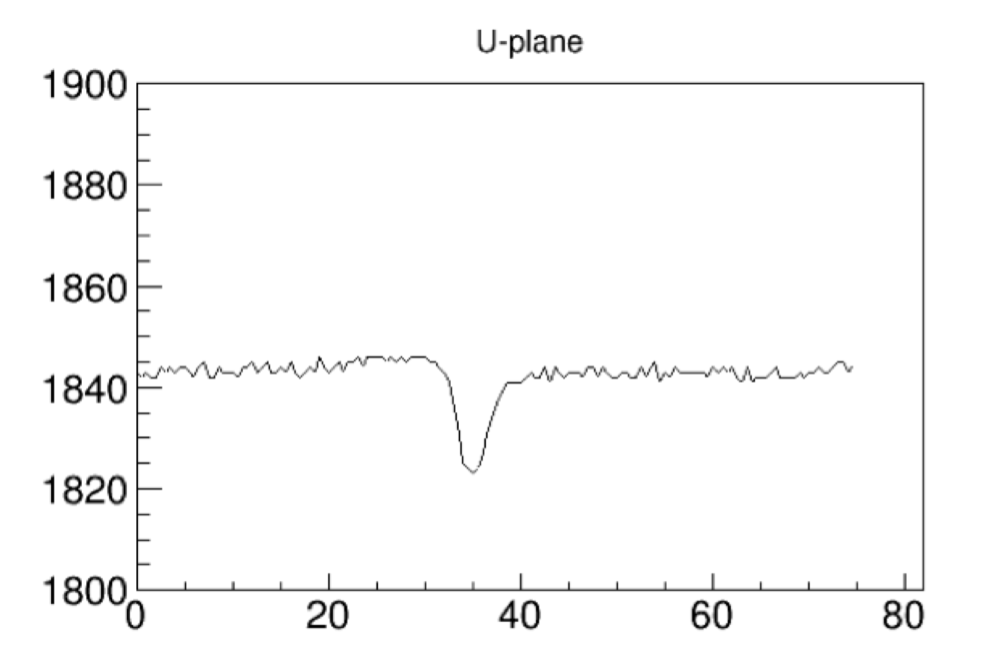}
\includegraphics[width=0.32\textwidth]{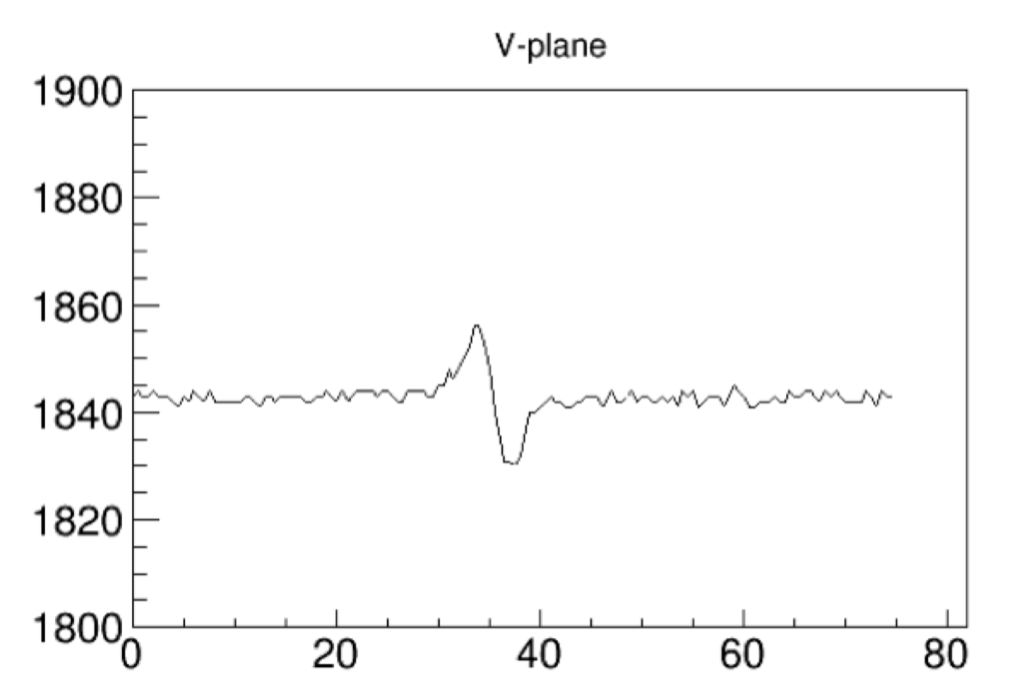}
\includegraphics[width=0.32\textwidth]{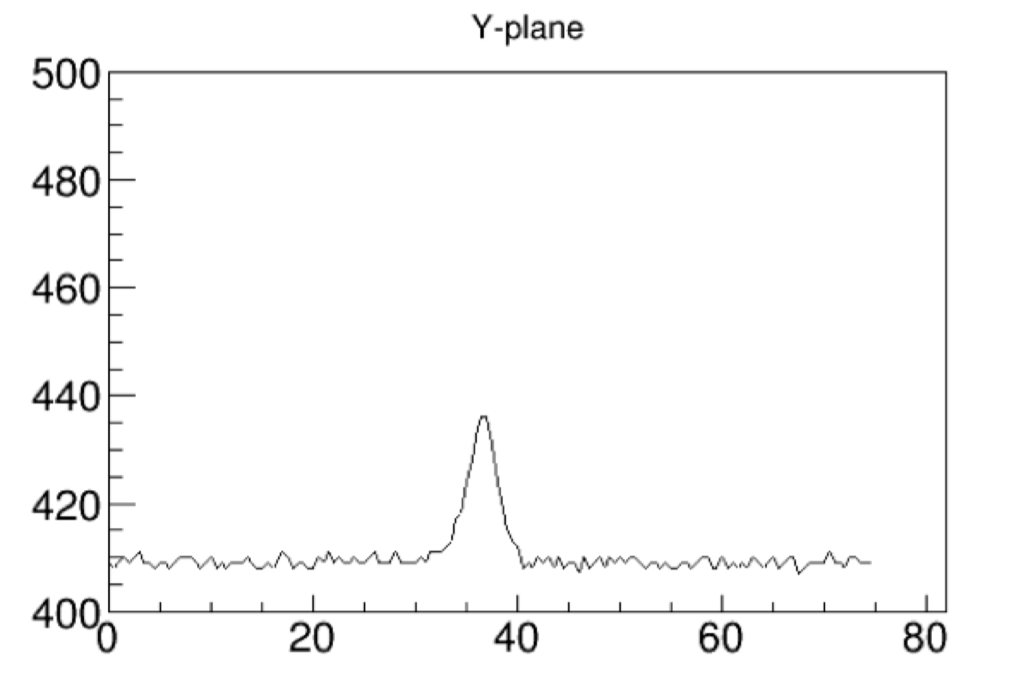}
\\
\includegraphics[width=0.32\textwidth]{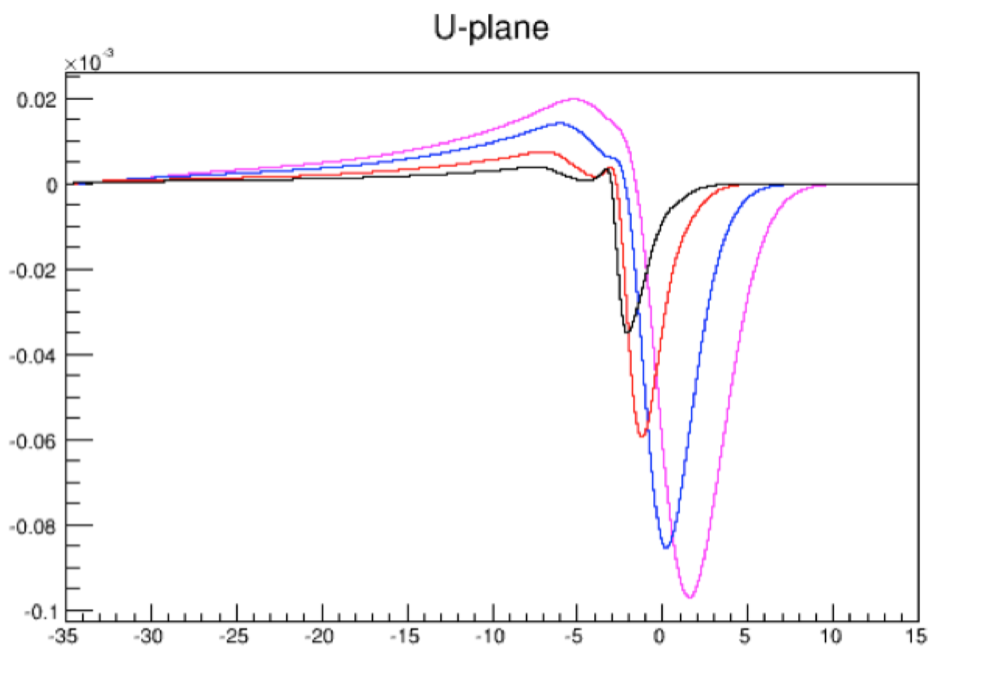}
\includegraphics[width=0.32\textwidth]{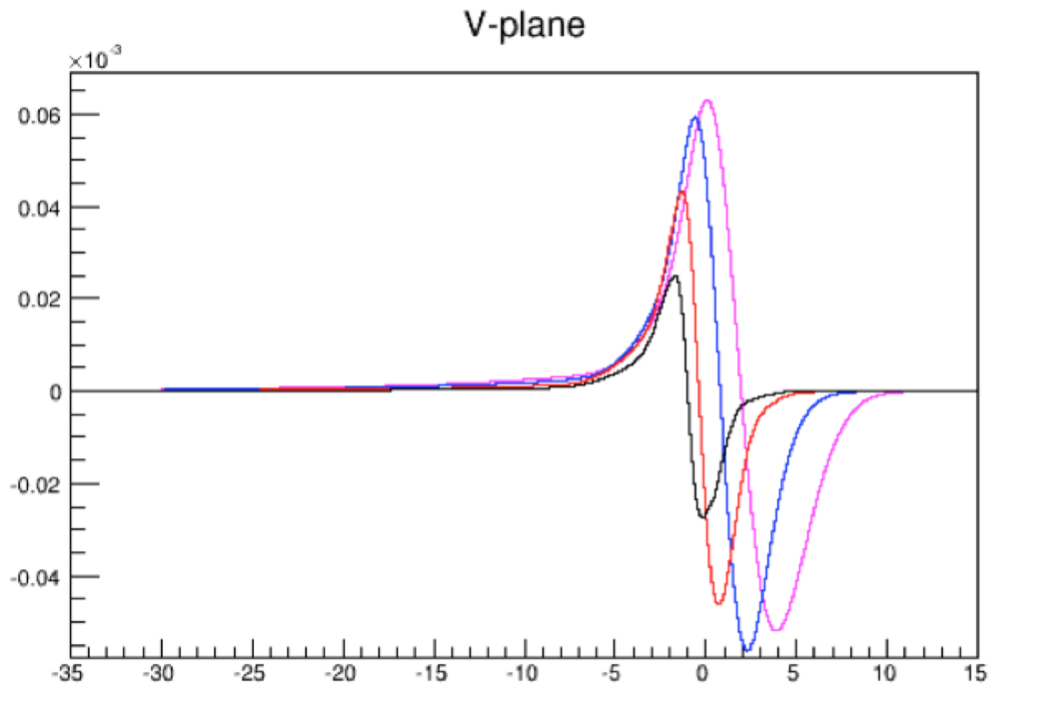}
\includegraphics[width=0.32\textwidth]{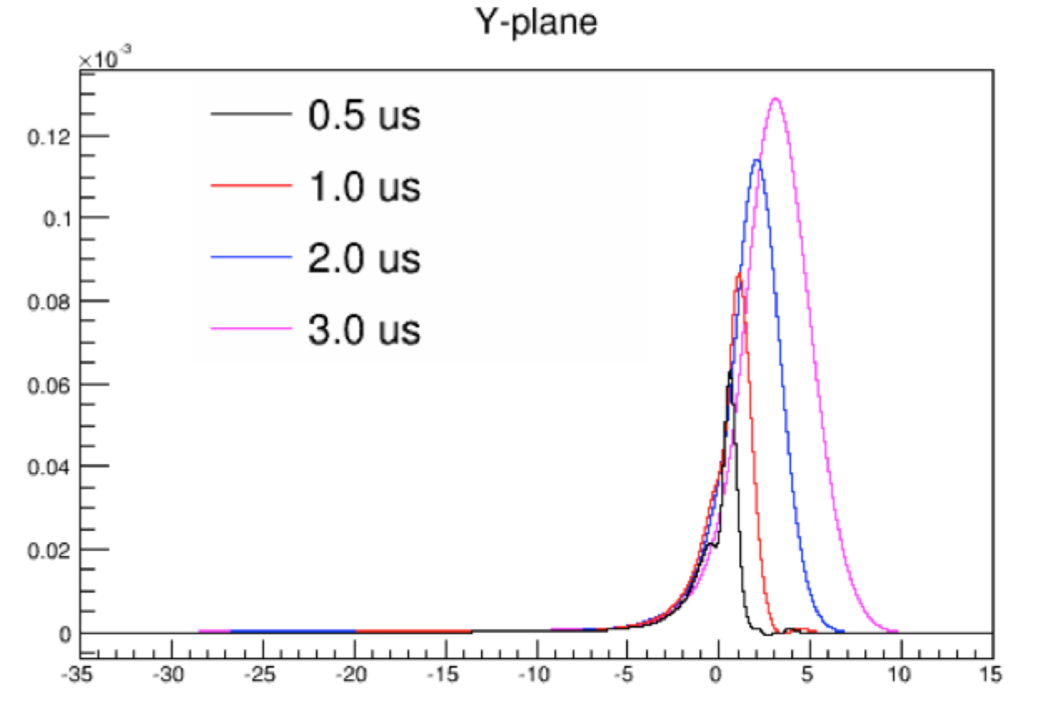}      
\caption[raw_conv]{Raw Signal (top row); Convoluted Signal (bottom row) for U, V and Y-Plane respectively}
\label{raw_conv}
\end{figure}

In this section the true raw wire signal and the actual signal obtained
after processing by the readout electronics are described. The
digitized signal obtained after the ADC is formed when the ionization
signal is convoluted with the detector and the front-end cold
electronics response functions and then digitized at a
fixed frequency. The
top row of Fig.~\ref{raw_conv} 
shows the raw MIP signal in the U, V and Y-Planes and in the bottom
row, the raw signal convoluted with field and 
electronics response for different shaping time settings are shown.     

\subsection{Noise Sources in Detector}
\label{sec:noise}

The readout electronics and digitization circuits are the two main sources of noise in the detector. 
In the case of the front-end readout electronics, the first  transistor noise is the main component~\cite{cold}. 
The first transistor noise contribution to the measured signal charge is proportional to the 
total capacitive load on the input channel, comprised of the sense wire capacitance, cable capacitance, 
and input transistor capacitance. This total capacitance limits the signal-to-noise 
ratio and it is the one dominant factor on which the feasibility and scalability of a 
LArTPC design critically depends. 
The other electronics noise sources are from thermal noise on the sense wires and 
connection leads (signal cable). Usually, thermal noise is made negligible by 
choosing appropriate resistors. Digitization noise arises during the
signal digitization by the ADC
which has a 12-bit resolution and a sampling rate of 2 mega-samples
per second (MS/s). The digitizer 
has been chosen in such a way to ensure that digitization noise is
much smaller than the front-end electronics noise. 

Other noise sources such as microphonics, and pick-up noise on the TPC
wires could also be present in the detector. These can be eliminated
using further signal processing steps that are not discussed in this
report.

\subsection{Deconvolution}
The next stage of signal processing is termed  ``Deconvolution", which means reversing 
the effects of convolution by unpacking and removing the readout electronics
and field response of the wire planes. The basic deconvolution process
is implemented in the standard LArSoft~\cite{larsoft} software signal
processing procedure which was originally developed by the ArgoNeuT
experiment~\cite{bruce}, 
and further developed by MicroBooNE. The process of deconvolution is
explained using the following equations. 

\begin{equation}
\label{eq:decon_1}
M(t_0) =  \int_{\{t\}} R(t-t_0) .\, S(t) \, dt  
\end{equation}

\begin{equation}
\label{eq:decon_2}
M(w)  =  R(w) . S(w) 
\end{equation}

\begin{equation}
\label{eq:decon_3}
S(w)  =   \frac {M(w)} {R(w)} 
\end{equation}

If M is the measured signal i.e, the digitized signal convoluted with
the response functions R, 
and S is the desired real signal, then the measured signal in the time domain is given by \Eq{decon_1}. 
In order to remove the effects of the different response functions, a 
fast fourier transformation (FFT)~\cite{fft} is 
performed on the measured signal in the time domain. The resulting
measured signal in the frequency domain is as shown in \Eq{decon_2}. By using simple factorization, 
the real deconvoluted signal (number of electrons reaching wire
planes) is then extracted in the 
frequency domain, \Eq{decon_3}. To obtain the real charge signal in time domain S(t), 
an inverse fourier transformation is performed. 

As discussed in the previous section, there are different types of noise sources present in 
the detector and in order to extract the true charge signal from the
measured signal, the noise contribution to the measured signal needs
to be eliminated as much as possible. The filtering of the noise from
the measured signal will be discussed in next section.

\subsection{Noise Filtering}
To remove noise from the deconvoluted signal, a Wiener noise filter
~\cite{wiener} is constructed using 
the expected signal and noise frequency response functions. The Wiener filter can be defined as 
in \Eq{filter}, where S is the signal and N is the noise in the frequency domain.

\begin{equation}
\label{eq:filter}
F(w)  =   \frac {S^2(w)} {S^2(w)+N^2(w)} 
\end{equation}

Wiener deconvolution is done in the frequency domain in order to minimize the impact of 
deconvolved noise at frequencies which have a poor signal-to-noise
ratio. Using the noise filter, 
$F(w)$, in the deconvolution method, \Eq{decon_3} is modified as such:

\begin{equation}
\label{eq:decon_filt}
S(w)  =   \frac {M(w)} {R(w)} . F(w)
\end{equation}

\begin{figure}[!h!tbp]
\includegraphics[width=0.24\textwidth]{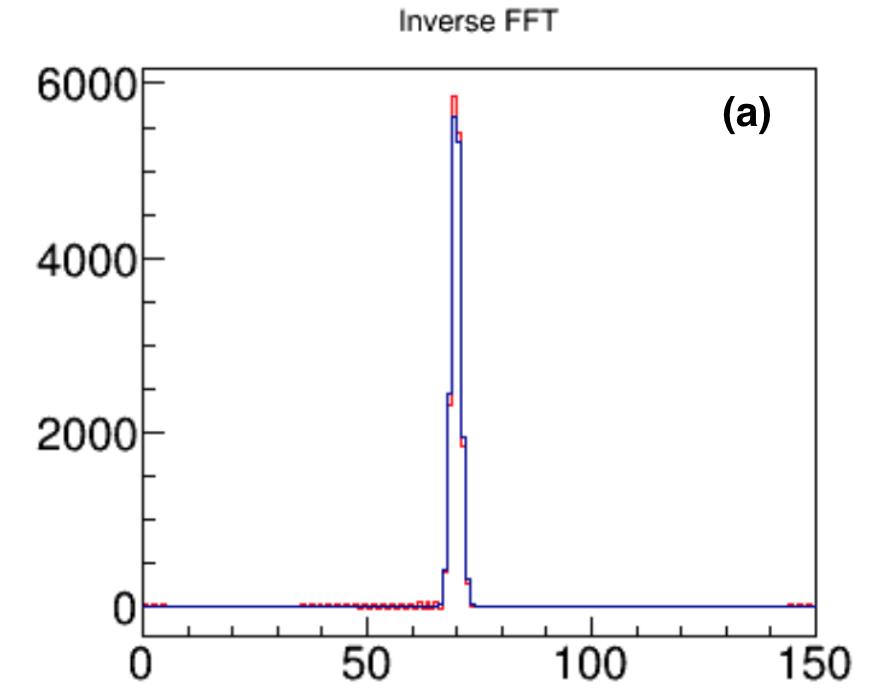}
\includegraphics[width=0.24\textwidth]{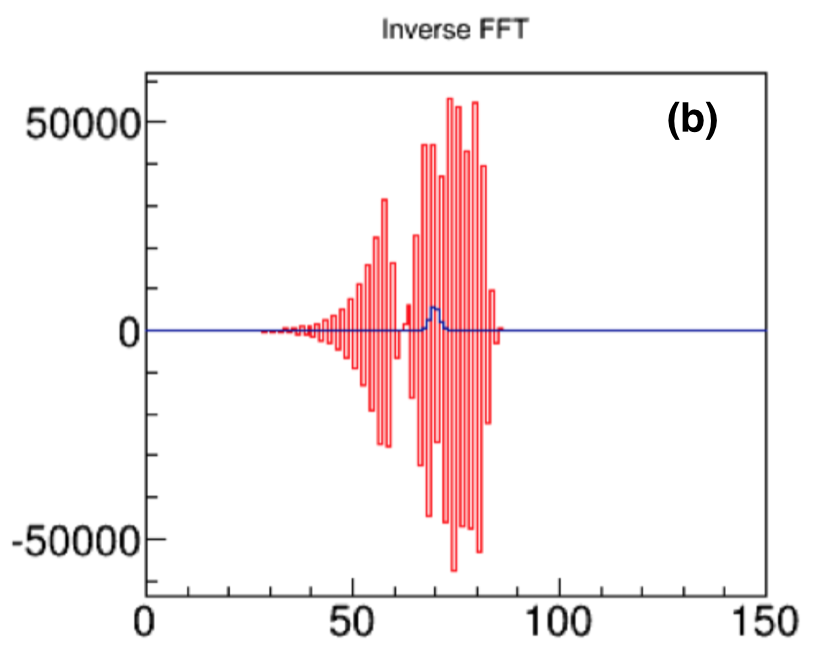}
\includegraphics[width=0.24\textwidth]{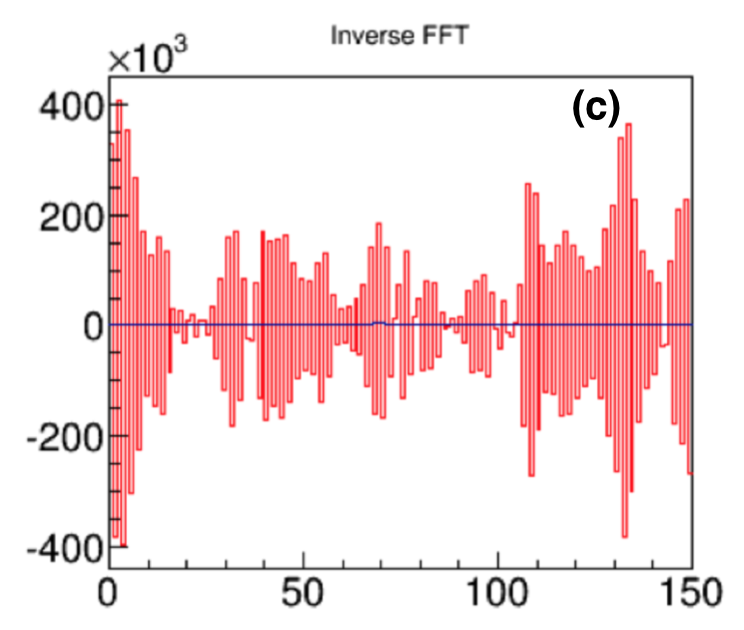}
\includegraphics[width=0.24\textwidth]{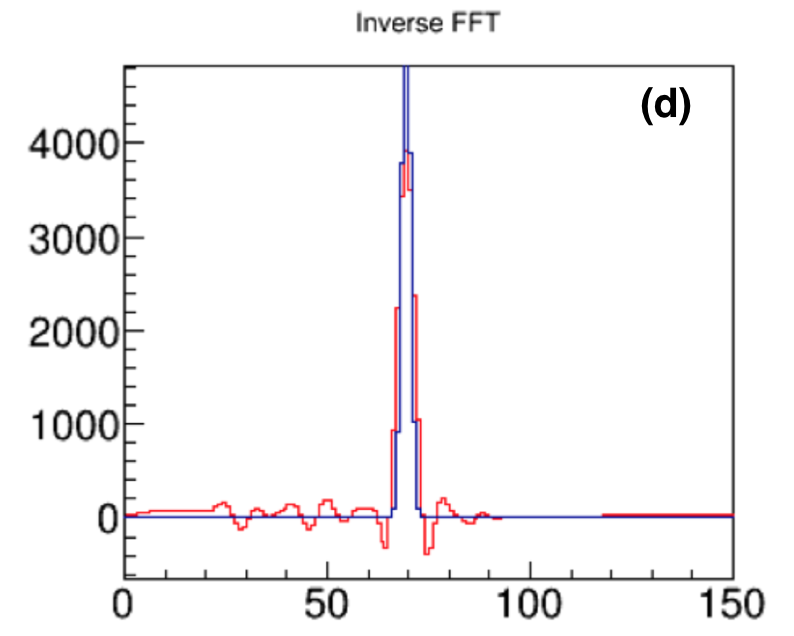}
\caption[noise_filt]{Effect of Signal Processing and Noise Filtering}
\label{noise_filt}
\end{figure}

The effect of noise and the process of noise filtering is described in Fig.~\ref{noise_filt} 
using toy Monte Carlo (MC) samples. Fig.~\ref{noise_filt} shows (a) a very good agreement of the true 
injected signal (blue) and the deconvoluted signal (red) without any noise in the time domain; 
(b) when the signal digitization is introduced, noise can be clearly seen in deconvoluted signal (red); 
(c) the deconvoluted signal  (red) after adding random white noise as an example of electronics 
noise where the deconvoluted signal peak can not be seen due to large amount of noise present; 
(d) after introducing the Wiener noise filter in the deconvolution step using \Eq{decon_filt}, 
the deconvoluted signal (red) is very close to the true injected signal (blue).      

The Wiener Filter function is optimized for different gain and shaping time settings for both induction 
and collection planes. The filter reduces the noise by a significant amount, however there is 
a loss in signal amplitude too. In order to preserve the signal
strength and to obtain the true charge information from all the three
planes, the filter is normalized in such a way so that it 
conserves the area. This step is very important in order to have a robust noise filter.

\section{Additional Signal Processing Challenges}

\begin{figure}[!h!tbp]
\includegraphics[width=0.48\textwidth]{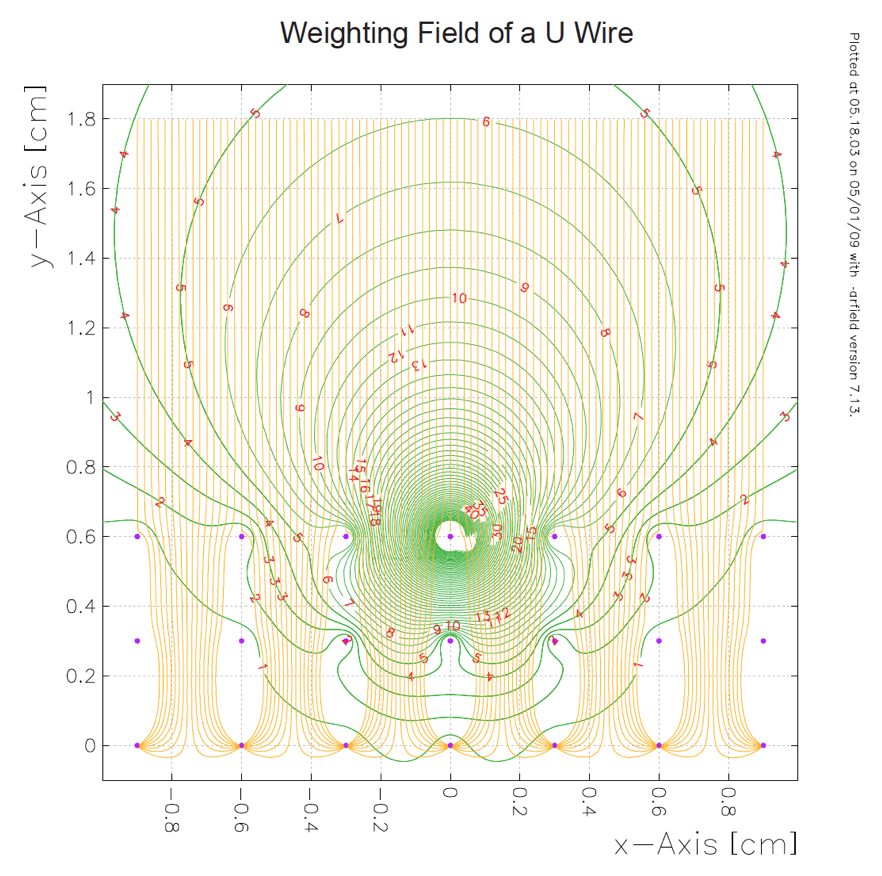} 
\includegraphics[width=0.48\textwidth]{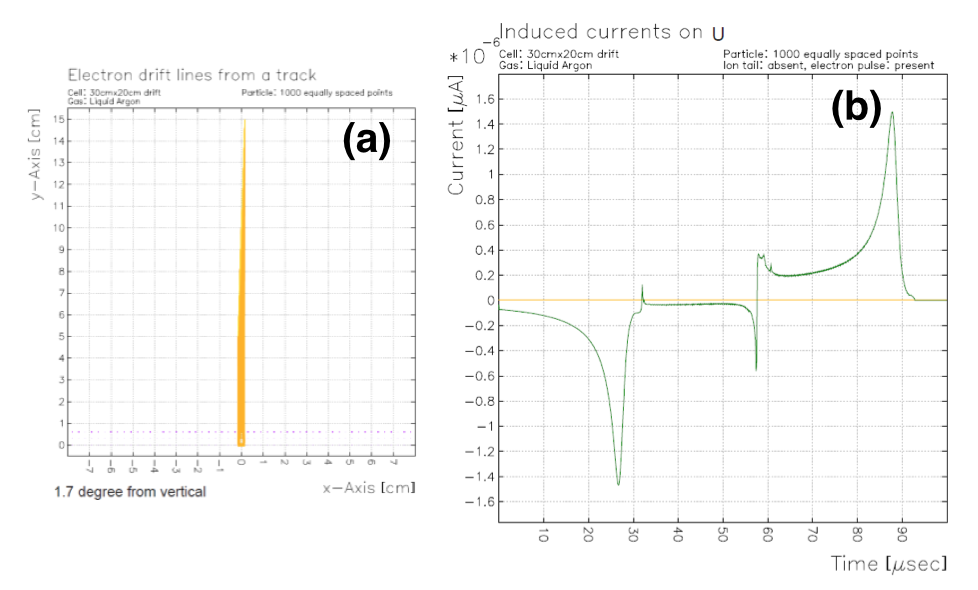}
\caption[dic]{2-D field simulations of the weighted potential distributions for a U-Plane wire. 
The weighed equipotential contours (green) are shown superimposed on
the electron drift lines shown in orange (left); and (a) for a track $1.7\degree$ from the vertical, (b) the induced current waveform obtained from the Garfield simulation on the central U-Plane wire.}
\label{dic}
\end{figure}

With the robust noise filtering and deconvolution, the full signal processing chain is complete 
and the desired charge signal is obtained from all three planes. There are still some additional challenges 
involved in the process. Due to the wire readout assembly of LArTPC, there is an effect of dynamic induced 
charge as ionized electrons traveling through the TPC wires induce signal not only on the closest 
wire but also on the adjacent wires. The field model described above does not take into account the 
charge contributions from the adjacent wires and treats signal from each TPC wire independently. 
To show the effect of induced current on the signal amplitude,
Fig.~\ref{dic} (left) shows the weighted 
equipotential contours (green) for a U-Plane wire superimposed on the
electron drift lines (orange). 
The induced charge on each wire is derived using \Eq{dic} along 
the drift line of each electron.

\begin{equation}
\label{eq:dic}
i  =    - q_m E_w . v_d
\end{equation} 
\\

Fig.~\ref{dic} (right) (b) shows the induced current waveform for a
central U-Plane wire using the 2-D GARFIELD 
simulation for a track $1.7\degree$ from the vertical (shown in
(a)). In comparison with the signal response 
in Fig.~\ref{resp} (left), the induced charge signal is more complicated and strongly depends on the angle of the track. 

In order to account for the dynamic induced charge effect, the
traditional deconvolution scheme described above is revised using the
2-D fast fourier transformation method in both time and wire parameter
space. This implementation of a double deconvolution method is the
most recent development in the LArSoft signal processing procedure. 

\section{New 3D Reconstruction with Charge and Time}
\begin{figure}[!h!tbp]
\includegraphics[width=0.48\textwidth]{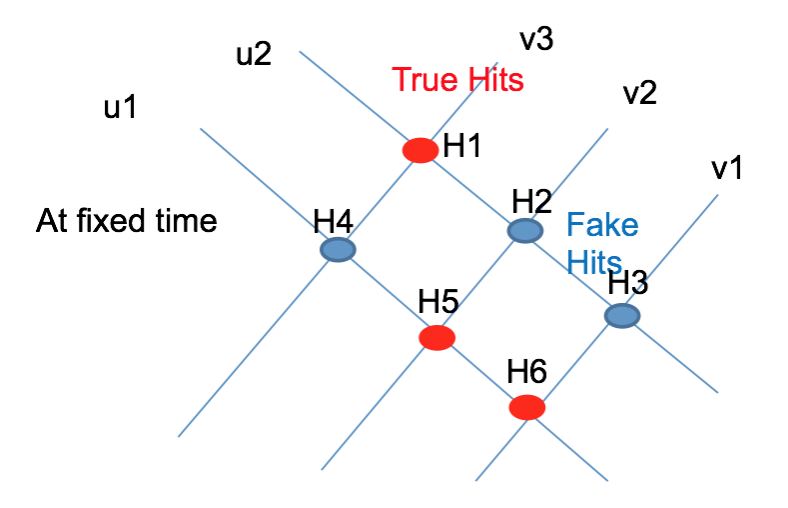} 
\includegraphics[width=0.48\textwidth]{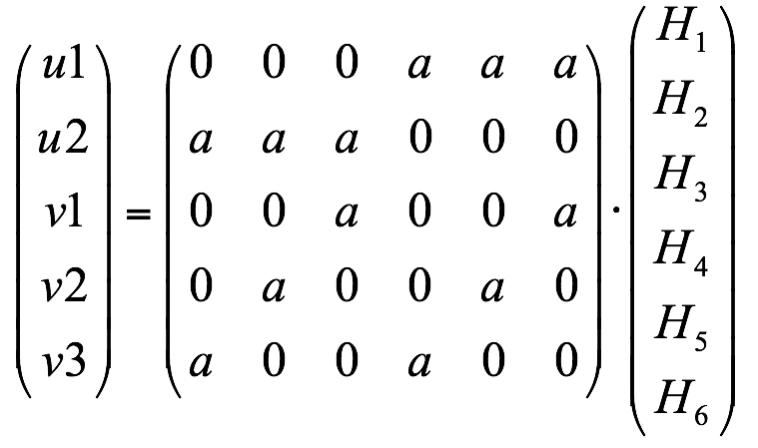}
\caption[wirecell]{An example of hit degeneracy using two planes (left); and charge matrix equation to solve degeneracy (right)}
\label{wirecell}
\end{figure}

After the robust signal processing chain, a new 3D event reconstruction method using both 
charge and time information is being developed. This new method is called ``Wire-Cell 
Reconstruction"~\cite{wirecell}. Reconstruction with a LArTPC wire
plane readout is challenging 
due to inherent ambiguities/degeneracies when using a projective wire geometry. In particular, the timing information alone is not enough 
to remove various ambiguities in a complex electromagnetic shower consisting of many tracks. 
An example of such a degeneracy is illustrated in Fig.~\ref{wirecell}
(left) using only two wire planes
for simplicity.  The true hits are shown in red and the fake hits in
blue. For a given time slice, a  
total of six possible hits on two U wires and three V wires generates ambiguation. This degeneracy 
increases exponentially with the increase in number of hits. Additional information is required in order to 
remove this degeneracy. 

Since the same charge is measured by all three wire planes, charge as
well as time information can be reliably used 
to resolve this degeneracy. Fig.~\ref{wirecell} (right) shows the charge matrix equations, where $u_i$, $v_i$ 
are the measured charges on the wires, $H_i$ matrix is the true charge to be resolved. After solving these 2D equations, the
charge on the fake hits is expected to be close to zero and hence the degeneracy is greatly reduced. This technique 
is used to obtain 3D hit maps by combining results from different time slices. This new algorithm is under rapid 
development towards the goal of automated reconstruction for LArTPC. 

\section{Conclusions}
The LArTPC is an excellent detector technology for precision neutrino
physics measurements. The MicroBooNE experiment, being the first
experiment in the future short
baseline program at Fermilab is an important step in the
development of LArTPC technologies
for future multi-kiloton detectors, in addition to enabling a wide range
of measurements of neutrino cross-sections and interactions. LArTPC
signal processing is the first and a critical
step in obtaining the correct charge and time information from all
three wire planes. After robust signal processing, we have
demonstrated that both the correct charge and
time information on all 3 planes can be obtained and is available
to be used in future improved 3-D tracking and calorimetry measurements.


\end{document}